\documentclass[12pt,a4paper]{article}

\usepackage[margin=2.5cm]{geometry}
\usepackage{graphicx}
\usepackage{amsmath}
\usepackage{amssymb}
\usepackage{bm}
\usepackage{authblk}
\usepackage{setspace}
\usepackage{lineno}
\usepackage{cmap}
\usepackage[T1]{fontenc}
\usepackage{xcolor}
\usepackage[colorlinks=true,
            linkcolor=black,
            urlcolor=blue,
            citecolor=blue,
            anchorcolor=blue]{hyperref}

\graphicspath{{./}{Figures/}}
\title{\bfseries What Photocurrent Versus Effective Voltage Tells Us About
Charge Generation in Organic Solar Cells}

\author[1,*]{Ardalan Armin}
\author[1]{Austin M.\ Kay}
\author[1]{Drew B.\ Riley}
\author[2]{Oskar J.\ Sandberg}
\author[1]{Paul Meredith}

\affil[1]{Centre for Integrative Semiconductor Materials, Department of Physics,
Swansea University, Swansea SA1 8EN, United Kingdom}
\affil[2]{Physics, Faculty of Science and Engineering, \AA bo Akademi University,
Turku, Finland}
\affil[*]{Corresponding author. E-mail: \texttt{ardalan.armin@swansea.ac.uk}}

\date{}

\begin{document}
\maketitle

\begin{abstract}
\noindent
The photocurrent of an organic solar cell is routinely plotted against an
effective voltage and used to extract an exciton dissociation probability, and
through it a free-charge generation efficiency. The photocurrent measures
neither. Using a drift-diffusion model in which free-charge generation is field
independent and losses occur only by non-geminate recombination, we show that the
apparent dissociation probability returned by this procedure lies well below
unity even for complete dissociation. Across wide variations of carrier mobility
and recombination strength it follows the fill factor and
coincides with the short-circuit charge collection efficiency, reporting neither
the generation yield nor anything about collection beyond what the fill factor
already shows. Even for ideal transport the normalised photocurrent saturates
below unity at an intensity-independent first-order limit set by recombination of
photogenerated carriers at the contacts. Applied to ten organic solar cells across four material classes whose generation efficiencies are known independently, the apparent dissociation probabilities track the short-circuit collection efficiency rather than the generation yield. We therefore suggest this construction not be used,
particularly in the high-efficiency non-fullerene acceptor systems that are now
state of the art.
\end{abstract}

\newpage
\section*{Introduction}
At any given light intensity, the power-conversion efficiency of an organic solar cell is determined by its short-circuit current, open-circuit voltage, and fill factor, and a recurring ambition is to decompose the current into two parts. The first part reflects how efficiently photons are converted into free charge carriers, and the second part reflects how efficiently free charge carriers are collected. A popular route to this decomposition starts from the photocurrent ($J_{\mathrm{ph}}$), which at any given voltage bias ($V$) is defined by the difference between the current density measured under illumination ($J_{\mathrm{light}}$) and the current density measured in the dark ($J_{\mathrm{dark}}$); i.e.,
\begin{equation}\label{eq:Jph}
J_{\mathrm{ph}}(V)=J_{\mathrm{light}}(V)-J_{\mathrm{dark}}(V).
\end{equation}
Often, the photocurrent is plotted against the effective voltage $V_{\mathrm{eff}}=V_{0}-V$, where $V_{0}$ is a compensation voltage defined by $J_{\mathrm{ph}}(V_0)\equiv0$. As the reverse bias is increased, the magnitude of $J_{\mathrm{ph}}$ rises before tending toward a saturation value ($J_{\mathrm{sat}}$).

This photocurrent versus effective voltage procedure, herein shortened to the $J_{\mathrm{ph}}$--$V_{\mathrm{eff}}$ procedure, was introduced by Mihailetchi et al. in 2004, who described the field and temperature dependence of $J_{\mathrm{ph}}$ in a polymer--fullerene blend~\cite{Mihailetchi2004}. The photo\-current is given by
\begin{equation}\label{eq:Jph2}
J_{\mathrm{ph}}(V)=P_{\mathrm{gen}}(V)P_{\mathrm{coll}}(V)J_{\mathrm{sat}},
\end{equation}
where $P_{\mathrm{gen}}(V)$ is the true dissociation efficiency of bound electron-hole pairs, written $P_{\mathrm{diss}}$ in the original work and renamed here because it is also the free-charge generation efficiency, and $P_{\mathrm{coll}}(V)$ is the collection efficiency of free charge carriers. In the work by Mihailetchi et al., $P_{\mathrm{gen}}(V)$ was described by the Onsager--Braun model of field-assisted dissociation and $P_{\mathrm{coll}}(V)$ by the Sokel--Hughes model which accounts for the loss due to diffusion \cite{Mihailetchi2004,Onsager1938,Braun1984,Sokel1982}. As the field dependence of geminate pair separation genuinely limited the photocurrent in their system, the $J_{\mathrm{ph}}$--$V_{\mathrm{eff}}$ curves could be fitted across a wide voltage range. From this point of inception, a simplified procedure became standard over the years, with more recent publications often using a single short-circuit ratio only \cite{Kyaw2013,Proctor_2013}. This more-recent version of the $J_{\mathrm{ph}}$--$V_{\mathrm{eff}}$ procedure is the focus of this work.

Nowadays, the $J_{\mathrm{ph}}$--$V_{\mathrm{eff}}$ procedure routinely applied in countless publications commonly takes the following form. the quantity reported as the dissociation probability, which we write $P_{\mathrm{diss}}^{\mathrm{app}}$ throughout to keep it distinct from the true yield $P_{\mathrm{gen}}$ of Equation~(\ref{eq:Jph2}), is quantified as the ratio of the photocurrent at short circuit to the saturation current:
\begin{equation}\label{eq:Pdiss}
P_{\mathrm{diss}}^{\mathrm{app}}=\frac{J_{\mathrm{ph}}(V=0)}{J_{\mathrm{sat,exp}}},
\end{equation}
where $J_{\mathrm{sat,exp}}=J_{\mathrm{ph}}(V=V_{\mathrm{rev}})$ is the saturation current estimated from a high reverse bias voltage $V_{\mathrm{rev}}$ (typically $V_{\mathrm{rev}}=-1$~V). Furthermore, the maximum exciton generation rate is read as $G_{\mathrm{ex}}=|J_{\mathrm{sat,exp}}|/qd$, with $q$ being the elementary charge and $d$ the active-layer thickness.

Over the years, Equation (\ref{eq:Pdiss}) migrated into a routine diagnostic in which $P_{\mathrm{diss}}^{\mathrm{app}}$ is now reported as the exciton dissociation probability and readily interpreted as a material-specific charge generation efficiency. This usage is now especially common in the synthesis-led non-fullerene-acceptor literature, where a value of $P_{\mathrm{diss}}^{\mathrm{app}}$ near unity is taken as evidence of efficient free-charge generation and a lower value as evidence of incomplete dissociation. The scale of this usage is readily quantified, with a manual survey identifying at least 249 papers published since January 2020, of which 234 appear in eight flagship journals of the field. These include 65 papers in Advanced Functional Materials, 61 in Advanced Materials, 44 in Nature Communications, and 39 in Advanced Energy Materials. Given that only these eight journals were surveyed systematically, the actual number of papers where the $J_{\mathrm{ph}}$--$V_{\mathrm{eff}}$ procedure was used is likely much higher. Moreover, in our survey, we find that the usage of this method is accelerating, appearing in 66 publications in 2024 compared to only 11 in 2020.  

Quantitatively extracting the free-charge generation yield has historically required specialised techniques. Time-delayed collection field measurements \cite{Kniepert2011, vandewal2014efficient, kurpiers2016dispersive} resolve the prompt extracted charge as a function of an applied collection field, but the method is an electro-optical pump-probe experiment demanding fast laser pulses, small-area devices (to limit RC effects), and a sensitive transient amplifier; consequently, these measurements are performed neither on typical devices nor under typical operation conditions. Integral-mode transient charge extraction \cite{Zeiske2022} relaxes the RC and amplifier constraints at the cost of a narrower voltage range. Alternatively, temperature-dependent internal quantum-efficiency spectroscopy \cite{TIQE} provides an indirect route in which the photocurrent and its high-reverse-bias saturation are fitted across temperature to separate generation from recombination. Each of these techniques is involved, requiring dedicated apparatus and careful analysis, and each exists precisely because the generation yield cannot simply be read off a steady-state current--voltage curve. It is therefore striking that the simple $J_{\mathrm{ph}}$--$V_{\mathrm{eff}}$ procedure is so often used in their place.

There are good reasons to doubt that the short-circuit ratio [Equation (\ref{eq:Pdiss})] measures the charge generation efficiency at all, and the principal ones are physical rather than procedural. Firstly, substituting Equation (\ref{eq:Jph2}) into Equation (\ref{eq:Pdiss}) shows what the ratio actually contains,
\begin{equation}\label{eq:ratio}
P_{\mathrm{diss}}^{\mathrm{app}}=\underbrace{\frac{P_{\mathrm{coll}}(V=0)}{P_{\mathrm{coll}}(V=V_{\mathrm{rev}})}}_{\text{collection ratio}}\times\underbrace{\frac{P_{\mathrm{gen}}(V=0)}{P_{\mathrm{gen}}(V=V_{\mathrm{rev}})}}_{\text{generation ratio}},
\end{equation}
Equation (\ref{eq:ratio}) is exact, and it makes the requirements explicit. The procedure recovers the generation yield only if the collection ratio is unity, that is if $P_{\mathrm{coll}}$ is independent of voltage over $V_{\mathrm{rev}}\leq V\leq0$, and if generation has saturated at the reverse bias, $P_{\mathrm{gen}}(V_{\mathrm{rev}})=1$. Conversely, if generation is field independent then the generation ratio is exactly unity whatever the value of $P_{\mathrm{gen}}$, and the ratio collapses to $P_{\mathrm{coll}}(0)/P_{\mathrm{coll}}(V_{\mathrm{rev}})$, which carries no information about generation at all. The first requirement is met in efficient inorganic cells, where the photocurrent is essentially flat across the range of operating voltages because collection occurs via diffusion and is therefore field independent.

Organic blends behave differently, however, as their free charge carriers are generally limited by drift under the influence of a built-in electric field. The low and frequently imbalanced mobilities of these charges make collection a voltage-dependent competition between extraction and non-geminate recombination, leading to the magnitude of $J_{\mathrm{ph}}$ rising gradually with reverse bias, and not necessarily reaching saturation across the operating window \cite{Neher2016,Wurfel2015,Sandberg2024}. As a result, Equation (\ref{eq:Pdiss}) measures how far this voltage-dependent collection has progressed, which is a transport property rather than the fraction of absorbed photons that have produced free charge. The same competition between extraction and recombination fixes the fill factor \cite{Bartesaghi2015}, which is why the two are not independent, as discussed shortly.

A second and lesser difficulty concerns the abscissa. Superposition generally fails under illumination, so the light and dark characteristics do not cross at the built-in voltage ($V_{\mathrm{bi}}$), and $V_{0}$ is therefore an unreliable estimate of $V_{\mathrm{bi}}$. This is further complicated in undoped films as the electric field strength across the bulk is governed by injected charge, and therefore deviates from the nominal $(V_{\mathrm{bi}}-V)/d$, with the internal voltage driving charge collection differing from the external bias on the axis \cite{Sandberg2024, Kirchartz2015}. The illuminated characteristic is in addition displaced from the implied one by the voltage drop associated with the transport resistance of the active layer \cite{Saladina2025}, so the crossing point that fixes $V_{0}$ reflects charge transport rather than the built-in potential.

In modern state-of-the-art non-fullerene blends, free charges are generated with near-unity yield ($P_{\mathrm{gen}}\simeq1$)~\cite{TIQE}, and as that yield approaches unity its residual field dependence vanishes, since in the Onsager--Braun form the field enters only through the geminate-loss fraction $1-P_{\mathrm{gen}}$. Therefore, the field-assisted dissociation that originally motivated the $J_{\mathrm{ph}}$--$V_{\mathrm{eff}}$ procedure is largely absent in precisely the materials to which it is now routinely applied. If generation is almost wholly-efficient and rather field-independent, as it is by construction in the model below, a sub-unity $P_{\mathrm{diss}}^{\mathrm{app}}$ could not be a dissociation yield; so what does this often-used procedure actually tell us about organic donor:acceptor blends, and what does the ratio given by Equation (\ref{eq:Pdiss}) actually measure?

In the remainder of this work, we answer that question directly. Instead of fitting experimental data with yet another model, we first apply the standard $J_{\mathrm{ph}}$--$V_{\mathrm{eff}}$ procedure to current-voltage curves simulated using a numerical drift-diffusion model, in which the free-charge generation efficiency is exactly unity (or otherwise known) and field independent. In other words, there are no charge generation losses; every deficit returned by the $J_{\mathrm{ph}}$--$V_{\mathrm{eff}}$ procedure must therefore be due to collection losses. We find that $P_{\mathrm{diss}}^{\mathrm{app}}$ extracted from a photocurrent curve using Equation (\ref{eq:Pdiss}) is, in fact, numerically equal to the short-circuit collection efficiency and, for a given device architecture, follows the fill factor. This $P_{\mathrm{diss}}^{\mathrm{app}}$ never approaches the true dissociation efficiency, regardless of whether a field-dependent dissociation rate is used or not, and even for arbitrarily-efficient charge transport it saturates at a first-order transport loss ceiling below unity. As a final investigation, we then apply the $J_{\mathrm{ph}}$--$V_{\mathrm{eff}}$ procedure to ten organic solar cells whose free-charge generation efficiencies have been measured independently, four of them high-efficiency blends we characterised previously by temperature-dependent internal quantum efficiency measurements and the remaining six drawn from the literature, before showing that the apparent dissociation efficiency tracks the fill factor and not the measured free charge generation efficiency. Any comparisons between the exciton dissociation efficiencies of different material systems made using the $J_{\mathrm{ph}}$--$V_{\mathrm{eff}}$ procedure are therefore difficult to justify.

\section*{Results}

\subsection*{Model system and the construction of the procedure}

\begin{figure}[t]
\centering
\includegraphics{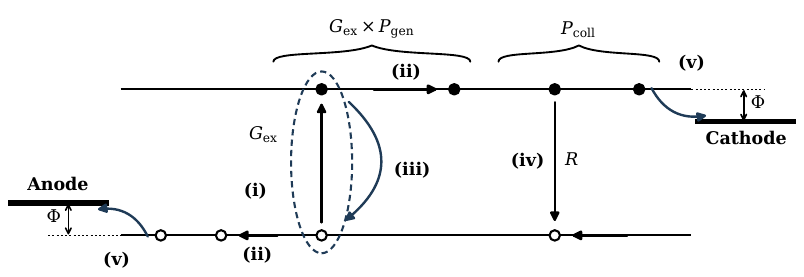}
\caption{\textbf{Schematic of the charge generation and collection process considered in this work.} The process begins with \textbf{(i)} photoexcitation and exciton formation, which is assumed to occur uniformly across the bulk at fixed rate $G_{\mathrm{ex}}$. Note that the bound electron-hole pair is encircled with a dashed line. A photogenerated exciton may either \textbf{(ii)} dissociate into a free electron-hole pair with efficiency $P_{\mathrm{gen}}$, the true dissociation probability and the quantity the $J_{\mathrm{ph}}$--$V_{\mathrm{eff}}$ procedure is intended to return, or \textbf{(iii)} undergo geminate recombination (where the electron recombines with the hole it produced upon photoexcitation). The combined free charge generation rate is given by the product $G_{\mathrm{ex}}P_{\mathrm{gen}}$. As these free charge carriers drift towards their respective contacts, they may undergo \textbf{(iv)} bimolecular (non-geminate) recombination in the bulk at a carrier density-dependent rate $R$ given by Equation (\ref{eq:BiRec}). Alternatively, free electrons and holes may be \textbf{(v)} collected at the cathode and anode, respectively, with both having symmetric injection barriers ($\Phi$). Note that recombination with injected carriers may also lead to losses. In total, the efficiency of the charge collection process is $P_{\mathrm{coll}}$.}
\label{fig:StateDiagram}
\end{figure}

We first apply the $J_{\mathrm{ph}}$--$V_{\mathrm{eff}}$ procedure to current--voltage
curves generated with a numerical drift-diffusion model in which the free-charge
generation efficiency is prescribed rather than inferred. The processes retained
in the model are summarised in Figure~\ref{fig:StateDiagram}. Excitons are
generated uniformly throughout the active layer at a rate $G_{\mathrm{ex}}$ and
dissociate into free electron-hole pairs with efficiency $P_{\mathrm{gen}}$, and
the resulting free carriers are either collected at the contacts, lost to
bimolecular recombination in the bulk, or lost to recombination with injected charge
at the electrodes. Full details of the model, the used parameters, and the
implementation of a field-dependent generation yield are given in Methods.

\subsection*{A model system with unity charge generation efficiency}

\begin{figure}[h]
\centering
\includegraphics[width=\textwidth]{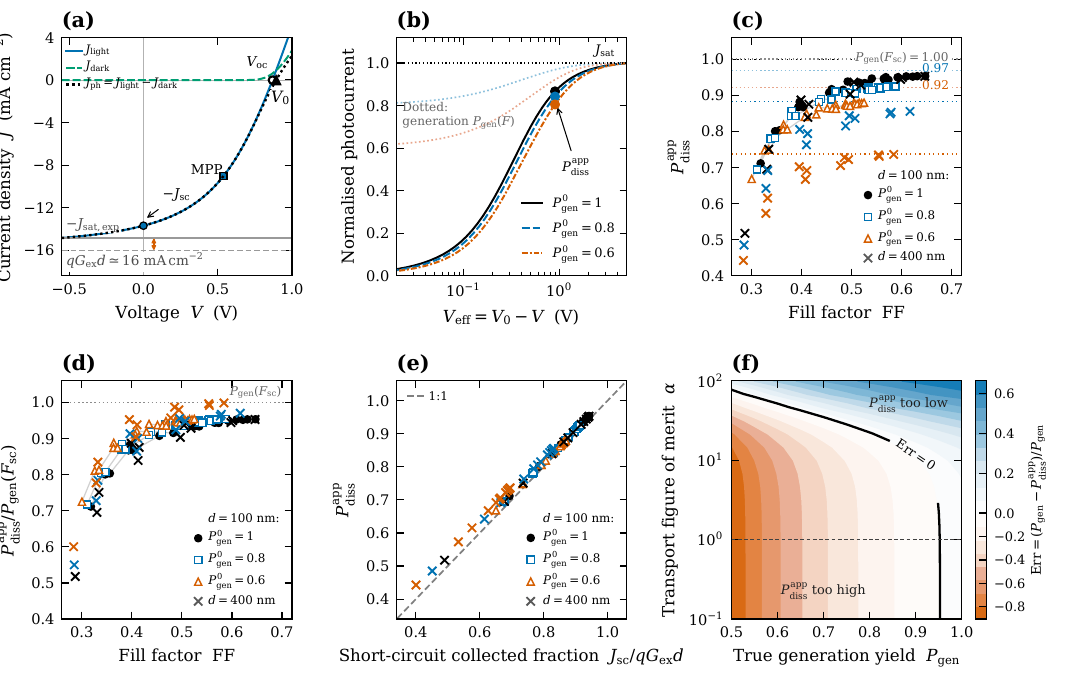}
\caption{\textbf{The $J_{\mathrm{ph}}$--$V_{\mathrm{eff}}$ procedure applied to simulated devices.} \textbf{(a)}~Construction of the $J_{\mathrm{ph}}$--$V_{\mathrm{eff}}$ procedure on a simulated device with uniform generation at $G_{\mathrm{ex}}=10^{22}\,\mathrm{cm}^{-3}\,\mathrm{s}^{-1}$, unity field-independent charge generation, balanced mobilities $\mu_{n}=\mu_{p}=2\times10^{-4}~\mathrm{cm^{2}\,V^{-1}\,s^{-1}}$, full Langevin recombination ($\gamma=1$), $d=100$~nm and $V_{\mathrm{bi}}=0.9$~V. \textbf{(b)} The photocurrent $J_{\mathrm{ph}}=J_{\mathrm{light}}-J_{\mathrm{dark}}$ is normalised to its reverse saturation $J_{\mathrm{sat}}$ and plotted versus effective voltage ($V_{\mathrm{eff}}$), with the $J_{\mathrm{ph}}$--$V_{\mathrm{eff}}$ procedure giving $P_{\mathrm{diss}}^{\mathrm{app}}=J_{\mathrm{ph}}(V=0)/J_{\mathrm{sat,exp}}$ at short circuit. Using an electric field $|F|=(V_{\mathrm{bi}}-V)/d$, the effect of a field-dependent generation efficiency was incorporated by scaling the exciton generation rate by the Onsager--Braun yield $P_{\mathrm{gen}}(F)$ given by Equation (\ref{eq:OnsagerBraunGen}), where zero-field values $P_{\mathrm{gen}}^0=1$, $0.8$, and $0.6$ were used (solid, dashed and dash-dotted curves); the corresponding $P_{\mathrm{gen}}(F)$ curves are dotted. \textbf{(c)} $P_{\mathrm{diss}}^{\mathrm{app}}$ against fill factor for three groups of zero-field generation efficiencies, each spanning mobilities from $5\times10^{-5}$ to $2\times10^{-3}~\mathrm{cm^{2}\,V^{-1}\,s^{-1}}$ and $\gamma=0.05$ to $1$ at fixed $V_{\mathrm{bi}}$ and $d$. Each group approaches its own short-circuit generation ceiling $P_{\mathrm{gen}}(F_{\mathrm{sc}})=1.00$, $0.97$, $0.92$ (dotted lines) without crossing it, the residual being collection loss. Crosses in \textbf{(c)}, \textbf{(d)} and \textbf{(e)} repeat the sweep for a $400$~nm active layer, with mobilities raised so that the fill factors overlap and the absorbed flux held at $qG_{\mathrm{ex}}d=16~\mathrm{mA\,cm^{-2}}$; the short-circuit field is four times weaker there, so the same zero-field yields give the lower ceilings $P_{\mathrm{gen}}(F_{\mathrm{sc}})=1.00$, $0.88$ and $0.74$ (dotted lines). \textbf{(d)}~The data of \textbf{(c)} divided by each device's own $P_{\mathrm{gen}}(F_{\mathrm{sc}})$, which collapses both thicknesses and all three generation families onto a single curve. \textbf{(e)} Plotted against the true short-circuit collected fraction $J_{\mathrm{sc}}/qG_{\mathrm{ex}}d$ the three groups collapse onto the line of unit slope, establishing that $P_{\mathrm{diss}}^{\mathrm{app}}=P_{\mathrm{gen}}(F_{\mathrm{sc}})\,P_{\mathrm{coll}}$; the small discrepancy above the line is due to $J_{\mathrm{sat,exp}}$ underestimating the true $J_{\mathrm{sat}}=-qG_{\mathrm{ex}}d$ at high reverse bias. \textbf{(f)}~Relative error of the procedure, $\mathrm{Err}=(P_{\mathrm{gen}}-P_{\mathrm{diss}}^{\mathrm{app}})/P_{\mathrm{gen}}$, against the transport figure of merit $\alpha$ [Equation~(\ref{eq:alpha})] and the true generation yield, for a constant field-independent $P_{\mathrm{gen}}$ with $J_{\mathrm{sat,exp}}$ read at $-4$~V. Solid line, $\mathrm{Err}=0$; dashed line, $\alpha=1$. Vermilion, overestimate of the generation yield; blue, underestimate. Source data are provided with this paper.}
\label{fig:recipe}
\end{figure}

To begin, the $J_{\mathrm{ph}}$--$V_{\mathrm{eff}}$ procedure was applied to a model device with unity charge generation efficiency; the results of this investigation are shown in Figure~\ref{fig:recipe}. The simulated current--voltage characteristics of this device are shown in panel~(a), with the photocurrent being given by the difference between the illuminated and dark branches and saturating toward $-J_{\mathrm{sat}}$ under reverse bias. Reading off the normalised photocurrent at short circuit in the usual way returns $P_{\mathrm{diss}}^{\mathrm{app}}=0.87$, which is well below unity despite the model containing no dissociation step whatsoever and no field dependence of generation; so the deficit is entirely due to losses in free carrier collection. Note that the choice of $V_0$ that shifts the effective voltage curve has no effect on the photocurrent read at short-circuit; it only shifts the curve left or right.

Figure \ref{fig:recipe}(b) relaxes the assumption of unity charge generation efficiency to show how a field dependence of generation affects the outcome of the $J_{\mathrm{ph}}$--$V_{\mathrm{eff}}$ procedure. The generation rate is scaled by the Onsager--Braun yield $P_{\mathrm{gen}}(F)$ given by Equation (\ref{eq:OnsagerBraunGen}), where a local electric field $|F|=(V_{\mathrm{bi}}-V)/d$ was assumed throughout the device. The yield was anchored to the zero-field values $P_{\mathrm{gen}}^0=1$, $0.8$, and $0.6$, while the transport was unchanged. Because $P_{\mathrm{gen}}(F)$ tends to unity at high reverse bias, the saturated current is common to all three, so the extracted $P_{\mathrm{diss}}^{\mathrm{app}}$ falls from $0.87$ to $0.84$ to $0.81$ with the short-circuit generation alone, nothing about extraction having changed. For each yield $P_{\mathrm{diss}}^{\mathrm{app}}$ value lies below the input $P_{\mathrm{gen}}(F)$, which are shown by the dotted lines for each zero-field value. This is true at the short-circuit field and at every field except for the high reverse bias-saturation limit, where both the generation and the collection efficiency approach unity, a regime of no relevance to an operating cell. These three numbers are Equation~(\ref{eq:ratio}) in action. Since $P_{\mathrm{gen}}$ has returned to unity at the reverse bias, the generation ratio is just $P_{\mathrm{gen}}(F_{\mathrm{sc}})=1.00$, $0.97$ and $0.92$; dividing each $P_{\mathrm{diss}}^{\mathrm{app}}$ by it leaves collection ratios of $0.868$, $0.870$ and $0.874$, the same to within $0.006$ across a $40$~\% change in the generation yield, as they must be when the transport is identical. The $J_{\mathrm{ph}}$--$V_{\mathrm{eff}}$ procedure therefore returns an apparent exciton dissociation probability as the product $P_{\mathrm{diss}}^{\mathrm{app}}\approx P_{\mathrm{gen}}(F_{\mathrm{sc}})\,P_{\mathrm{coll}}$ of the short-circuit generation yield and a transport-set charge collection efficiency $P_{\mathrm{coll}}\approx0.87$. This indicates that the addition of a field dependence of carrier generation results in the same misunderstanding as collection losses.

\subsection*{The apparent dissociation efficiency is the short-circuit collection efficiency}

To further substantiate the direct proportionality between charge collection and the $P_{\mathrm{diss}}^{\mathrm{app}}$ extracted using the $J_{\mathrm{ph}}$--$V_{\mathrm{eff}}$ procedure, we vary the carrier mobilities and the recombination reduction factor over wide ranges for the same zero-field charge generation efficiencies considered in Figure \ref{fig:recipe}(b). For each of these groups, Figure~\ref{fig:recipe}(c) demonstrates how $P_{\mathrm{diss}}^{\mathrm{app}}$ climbs with the fill factor toward a ceiling equal to the short-circuit generation yield for that zero-field generation rate; i.e., $P_{\mathrm{gen}}(F_{\mathrm{sc}})=1.00$, $0.97$ and $0.92$. The $P_{\mathrm{diss}}^{\mathrm{app}}$ never crosses it; the residual below the ceiling is the collection loss, fixed by the competition between extraction and recombination that also controls the fill factor \cite{Bartesaghi2015}. At low fill factors (i.e., for devices with lower carrier mobilities and/or high recombination rates), differences in charge-generation efficiency are negligible compared to losses in transport; data points for all three groups of generation rates therefore converge. In other words, the apparent dissociation efficiencies extracted for one blend with 100~\% charge-generation efficiency and another blend with 60~\% charge-generation efficiency are indistinguishable in the case that both blends have poor charge collection.

The correlation is not restricted to the $100$~nm devices. Repeating the sweep for a $400$~nm active layer, with the mobilities raised so that the fill factors overlap, leaves the $P_{\mathrm{gen}}^{0}=1$ family on the same locus, the mean deviation being $-0.013$. The two field-dependent families lie systematically lower, by $0.08$ and $0.15$ on average, but for a physical reason rather than a failure of the correlation: the short-circuit field falls from $9.0\times10^{6}$ to $2.25\times10^{6}~\mathrm{V\,m^{-1}}$ on going from $100$ to $400$~nm, so the same material has a genuinely lower Onsager--Braun yield at short circuit, $P_{\mathrm{gen}}(F_{\mathrm{sc}})=0.88$ and $0.74$ in place of $0.97$ and $0.92$. Dividing each point by its own ceiling collapses every family and both thicknesses onto one curve [Figure~\ref{fig:recipe}(d)], the mean deviations becoming $-0.013$, $+0.002$ and $+0.015$. A practical consequence follows: because the built-in field weakens as the absorber is made thicker, for a given blend the apparent dissociation probability will decrease with increasing active-layer thickness even though nothing about the material has changed.

These findings are consistent with the behaviour of a collection efficiency modulated by generation, not of a generation yield. The identification is made explicit in Figure~\ref{fig:recipe}(e), where $P_{\mathrm{diss}}^{\mathrm{app}}$ extracted using the $J_{\mathrm{ph}}$--$V_{\mathrm{eff}}$ procedure is plotted against the true short-circuit collected fraction $J_{\mathrm{sc}}/qG_{\mathrm{ex}}d$. Irrespective of the zero-field generation rate, all results follow the line of unit slope. The quantity the literature reports as an exciton dissociation probability is, to within the small offset between $J_{\mathrm{sat,exp}}$ and $qG_{\mathrm{ex}}d$, the charge collection efficiency at short circuit. Provided that $J_{\mathrm{sat,exp}}$ is evaluated under strong enough reverse bias, such that $J_{\mathrm{sat,exp}}\approx -qG_{\mathrm{ex}}d$, $P_{\mathrm{diss}}^{\mathrm{app}}\simeq J_{\mathrm{sc}}/qG_{\mathrm{ex}}d$ [Equation (\ref{eq:Pdiss})] is just the collection efficiency at short-circuit conditions, not the exciton dissociation yield.

Whether this collection-dominated quantity lies above or below the true generation yield depends on the transport quality. Figure~\ref{fig:recipe}(f) maps the relative error $\mathrm{Err}=(P_{\mathrm{gen}}-P_{\mathrm{diss}}^{\mathrm{app}})/P_{\mathrm{gen}}$ against the transport figure of merit introduced by Neher et al. \cite{Neher2016},
\begin{equation}\label{eq:alpha}
\alpha^{2}=\frac{q^{2}k_{2}G\,d^{4}}{4\mu_{n}\mu_{p}(k_{\mathrm{B}}T)^{2}},
\end{equation}
which combines the bimolecular recombination coefficient $k_{2}=\gamma\beta_{\mathrm{L}}$, the free-charge generation rate $G$, the active layer thickness and the two mobilities into a single parameter, with $\alpha<1$ corresponding to efficient charge transport and $\alpha>1$ to a transport-limited photocurrent. Herein, $\beta_{\mathrm{L}}=q(\mu_{n}+\mu_{p})/\varepsilon_{\mathrm{r}}\varepsilon_{0}$ is the Langevin encounter rate. The error changes sign across this plane, and it does so in a way that exposes the cancellation directly. In the ideal-transport limit $P_{\mathrm{diss}}^{\mathrm{app}}$ settles on its first-order ceiling of $0.953$ for every true yield in the range simulated, from $0.5$ to $1$: the quantity returned by the procedure has become completely independent of the generation efficiency it is supposed to measure. The procedure therefore underestimates a yield above that ceiling and overestimates any yield below it, and as transport degrades the apparent value falls further, until the error is positive at every yield. The sign is fixed jointly by the transport quality and by the true yield --- the $\mathrm{Err}=0$ contour moves from $\alpha\simeq2$ at $P_{\mathrm{gen}}=0.95$ to $\alpha\simeq76$ at $P_{\mathrm{gen}}=0.5$ --- so neither the size of the error nor even its direction can be inferred from the reported number.

\subsection*{A first-order ceiling on charge collection efficiency}

\begin{figure}[t]
\centering
\includegraphics[width=\textwidth]{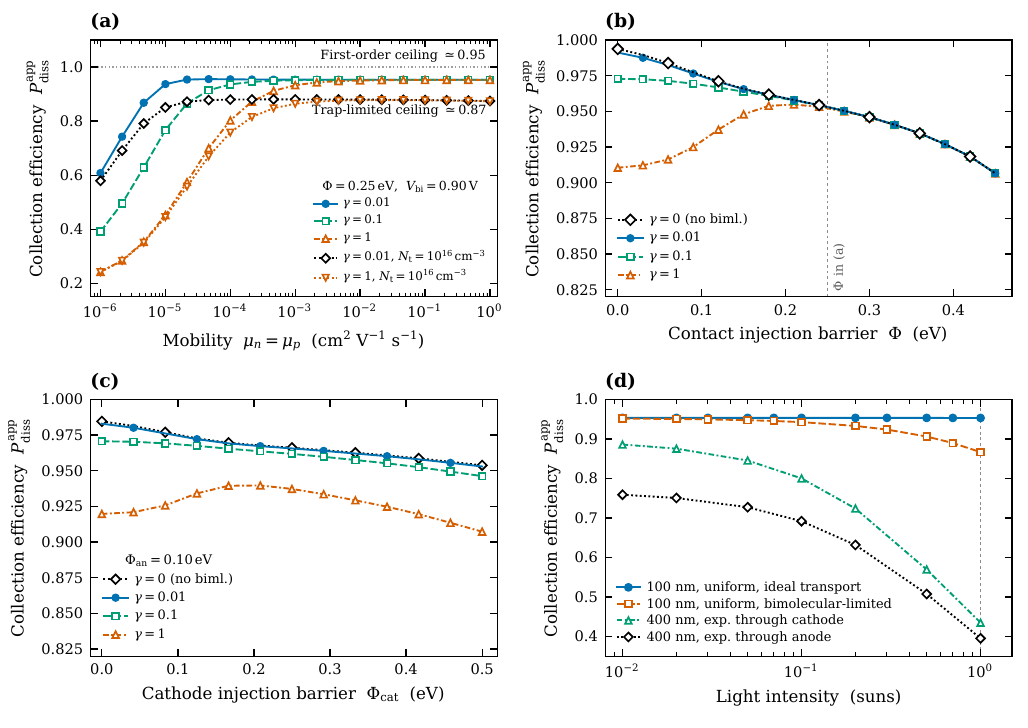}
\caption{\textbf{Exploring the first-order ceiling imposed on the collection efficiency.} \textbf{(a)}~The apparent dissociation probability ($P_{\mathrm{diss}}^{\mathrm{app}}$), or, more appropriately, the short-circuit collection efficiency, plotted against carrier mobility $\mu_{n}=\mu_{p}$ for symmetric injection barriers with $\Phi=0.25$~eV and three bimolecular reduction factors, and, for the two extreme reduction factors, with midgap traps added at a density $N_{\mathrm{t}}=10^{16}~\mathrm{cm^{-3}}$ (dotted, open diamonds and inverted triangles). Trap capture coefficients are taken from the Langevin encounter rate, $c_{n}=q\mu_{n}/\varepsilon_{\mathrm{r}}\varepsilon_{0}$ and $c_{p}=q\mu_{p}/\varepsilon_{\mathrm{r}}\varepsilon_{0}$, so that $c_{n}+c_{p}=\beta_{\mathrm{L}}$. \textbf{(b)}~Dependence of the ceiling on the contact injection barrier ($\Phi$) when transport is nearly ideal ($\mu=1~\mathrm{cm^{2}\,V^{-1}\,s^{-1}}$), the built-in voltage following $qV_{\mathrm{bi}}=E_{\mathrm{g}}-2\Phi$. \textbf{(c)}~The same sweep with asymmetric contacts: the anode barrier is held at $\Phi_{\mathrm{an}}=0.10$~eV while the cathode barrier is varied, so that $qV_{\mathrm{bi}}=E_{\mathrm{g}}-\Phi_{\mathrm{an}}-\Phi_{\mathrm{cat}}$. \textbf{(d)}~Intensity dependence of the charge collection efficiency. For the $100$~nm devices with uniform generation and $\Phi=0.25$~eV it is intensity-independent for ideal transport (the signature of a first-order loss) but increases with decreasing intensity in a bimolecular recombination-limited device (the signature of a second-order bulk process). Also shown is a $400$~nm device with asymmetric contacts ($0.10/0.40$~eV) and an exponential generation profile of absorption length $100$~nm entering through either contact, with the total absorbed flux equal to that of the $100$~nm reference. Mobilities are balanced and charge generation is unity and field independent in every panel. Source data are provided with this paper.}
\label{fig:ceiling}
\end{figure}

The collection efficiency can be affected by two main loss processes in the device: (i) first-order recombination (e.g. with injected or trapped charge carriers), and (ii) second-order, bimolecular recombination between photo-generated electrons and holes. If $P_{\mathrm{diss}}^{\mathrm{app}}$ returned by the $J_{\mathrm{ph}}$--$V_{\mathrm{eff}}$ procedure were limited only by second-order recombination, its shortfall from unity would vanish as the transport is improved. As explored in Figure \ref{fig:ceiling}, however, this is not the case. Figure~\ref{fig:ceiling}(a) shows the apparent exciton dissociation probability (or, more accurately, the short-circuit collection efficiency) versus the carrier mobility used in the simulation for three recombination strengths $\gamma=0.01$, $0.1$ and $1$, with generation complete and field-independent throughout. Note that symmetric injection barriers with $\Phi=0.25$~eV were used in these simulations. At low mobilities the collection is transport-limited and the three branches separate, with an increased $\gamma$ leading to higher collection losses and a reduced $P_{\mathrm{diss}}^{\mathrm{app}}$ being extracted. As the mobility increases the apparent dissociation efficiencies rise before plateauing at first-order ceilings, remaining invariant to orders-of-magnitude increase in mobility. Importantly, even as the bimolecular recombination rate is reduced to near-zero, neither faster transport nor weaker recombination lifts $P_{\mathrm{diss}}^{\mathrm{app}}$ to unity. 

The situation is further exacerbated when including trap-assisted Shockley--Read--Hall (SRH) recombination in the bulk. Adding midgap trap states of density $N_{\mathrm{t}}=10^{16}~\mathrm{cm^{-3}}$, with capture coefficients taken from the Langevin encounter rate so that $c_{n}+c_{p}=\beta_{\mathrm{L}}$, lowers the plateau from $0.95$ to $0.87$ while leaving its mobility-independence intact. That invariance is not accidental: the capture coefficient scales with mobility exactly as the extraction rate does, so improving transport cannot outrun trap capture when the capture is itself transport-limited. The SRH recombination channel is first order in the midgap limit, $R\simeq\tfrac{c_nc_p}{(c_n+c_p)}N_{\mathrm{t}}n$ for $n\simeq p$, and the extracted probability is correspondingly insensitive to illumination, moving only from $0.862$ to $0.874$ between $0.01$ and one sun. Trap-assisted recombination therefore adds a further first-order channel alongside recombination with injected charge carriers, and lowers the ceiling further rather than relieving it.

Even in the absence of traps, a finite collection efficiency loss remains in the limit of high mobilities. In this limit, bimolecular recombination in the bulk is outrun by extraction and contributes negligibly to collection losses. The remaining collection loss is induced by additional first-order recombination pathways related to the two contacts, and its effect is explored further in Figure \ref{fig:ceiling}(b), the charge collection efficiency varying with the injection barrier height ($\Phi$). At small $\Phi$, the collection efficiency depends on $\gamma$, and the loss is dominated by the recombination of photo-generated carriers with injected charge carriers  \cite{Sandberg2024}. This same phenomenon sets the chemical capacitance of a metal--insulator--metal device \cite{Sandberg2026}. As the injected carrier densities scale as $\exp(-\Phi/k_{\mathrm{B}}T)$, the effect of injected charge carriers is reduced as $\Phi$ is increased. For large  $\Phi$ only the diffusion loss due to surface recombination at the wrong electrodes survives. This recombination channel is present even in the $\gamma=0$ limit and follows the Sokel--Hughes form,  $P_{\mathrm{coll}}=\coth(qV_{\mathrm{bi}}/2k_{\mathrm{B}}T)-2k_{\mathrm{B}}T/qV_{\mathrm{bi}}$, with the loss growing as the injection barrier is raised and the built-in field weakened. In the $\gamma=1$ limit, the injected-charge recombination dominates for low barriers, but is overtaken by the diffusion-induced surface recombination near $\Phi\simeq0.2$~eV, producing a maximum.

Neither the ceiling nor its first-order origin depends on the contacts being symmetric. Figure~\ref{fig:ceiling}(c) repeats the barrier sweep with the anode barrier held at $0.10$~eV while the cathode barrier is varied from $0$ to $0.5$~eV. The behaviour of Figure~\ref{fig:ceiling}(b) is reproduced in full: the collection efficiency falls monotonically with barrier height for weak bimolecular recombination, while for $\gamma=1$ the competition between recombination with injected charge at low barriers and diffusion-induced surface recombination at high ones again produces a maximum, here near $\Phi_{\mathrm{cat}}\simeq0.2$~eV. The $P_{\mathrm{diss}}^{\mathrm{app}}$ remains below unity everywhere, reaching at most $0.985$.

As a final verification that recombination with injected carriers is a first-order loss mechanism fixed by the electrodes, which, unlike bimolecular recombination between two photogenerated populations, is independent of illumination, intensity-dependent simulations were carried out, with the results shown in Figure~\ref{fig:ceiling}(d). For a device with ideal transport, the collection efficiency is independent of illumination intensity across two orders-of-magnitude, the signature of a first-order process. Alternatively, for a bimolecular-limited device the collection efficiency rises as the intensity is lowered, the signature of a second-order process, and approaches the same ceiling only in the low-intensity limit where the carrier densities are sufficiently small for bulk recombination to become negligible. This channel does not vanish with increased charge generation efficiency and has been identified as the dominant fill-factor loss in optimised organic cells \cite{Wurfel2019}. Relaxing the assumption of uniform photogeneration works in the same direction. Figure~\ref{fig:ceiling}(d) also shows a $400$~nm device carrying an exponential generation profile with an absorption length of $100$~nm and illuminated through either contact, with the total absorbed flux held equal to that of the $100$~nm reference. The $P_{\mathrm{diss}}^{\mathrm{app}}$ falls to $0.40$--$0.43$ at one sun, and its value depends on which face the light enters even though the material, the contacts and the photon count are identical. Both curves rise steeply as the intensity is lowered, as expected of a bimolecular-limited device, but they do not converge; the residual gap reflects the different first-order losses on the two sides. A thicker absorber with a realistic absorption profile therefore returns an apparent dissociation probability far below unity, and one that depends on the optical geometry of the measurement.

The consequence is that the $J_{\mathrm{ph}}$--$V_{\mathrm{eff}}$ procedure misreports the dissociation yield; even for devices with excellent transport that are measured at low light intensities. What this procedure returns, as Figure~\ref{fig:ceiling} makes plain, is the charge collection efficiency at short circuit, which is capped by the injection barriers at the contacts. The loss can be reduced by lowering the injection barriers, which raises the built-in voltage and the short-circuit field, but it cannot be eliminated \cite{Sandberg2024}. This is a significant reason the fill factors of organic solar cells fall below those of semiconductors such as the metal-halide perovskites, which under normal operating conditions have branching fractions that are predominantly non-excitonic \cite{lin2015electro}.

\subsection*{Experimental verification}

\begin{figure}[t]
\centering
\includegraphics{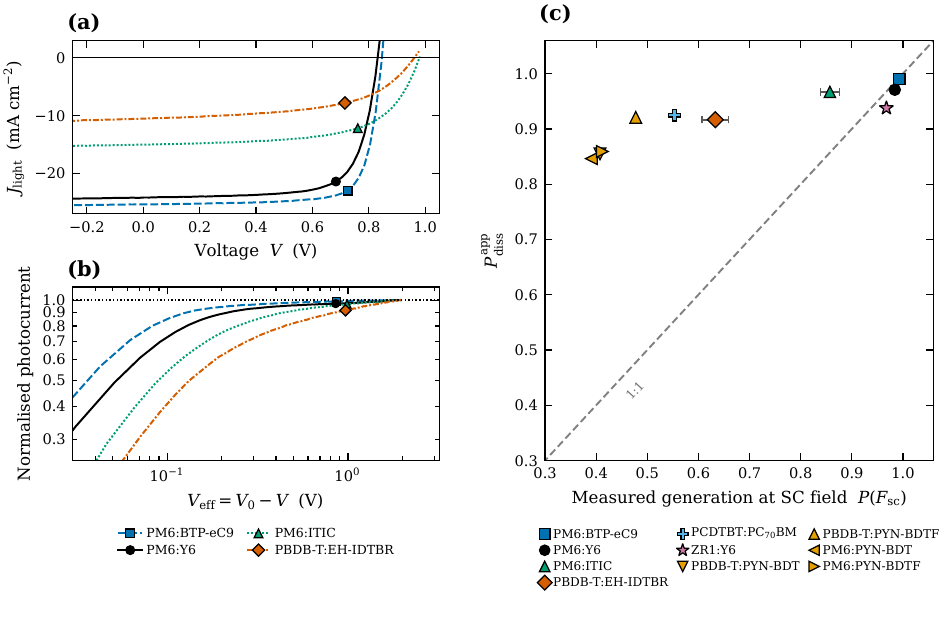}
\caption{\textbf{Application of the $J_{\mathrm{ph}}$--$V_{\mathrm{eff}}$ procedure to ten organic solar cells whose free-charge generation efficiencies have been measured independently.} \textbf{(a)}~Light current--voltage characteristics of the four polymer:non-fullerene blends of Ref.~\cite{TIQE}, with their maximum-power points marked (PBDB-T:EH-IDTBR has $d=90$~nm, the other three $d=100$~nm). \textbf{(b)}~Photocurrents of the same four blends, normalised to the measured saturation value ($J_{\mathrm{sat}}$). The effective voltages at which $P_{\mathrm{diss}}^{\mathrm{app}}$ values are evaluated using Equation~(\ref{eq:Pdiss}) are indicated by the symbols. Panels \textbf{(a)} and \textbf{(b)} show those four blends; locating $V_{0}$ for the six literature systems is not possible from their published characteristics, so all ten are compared at short circuit in \textbf{(c)}. \textbf{(c)}~$P_{\mathrm{diss}}^{\mathrm{app}}$ against the independently measured generation yield at the short-circuit field, for all ten systems. Data are from Ref.~\cite{TIQE} for PM6:BTP-eC9, PM6:Y6, PM6:ITIC and PBDB-T:EH-IDTBR; Ref.~\cite{Zeiske2022} for PCDTBT:PC$_{70}$BM; Ref.~\cite{Gerber2024} for ZR1:Y6; and Ref.~\cite{Alam2026} for the four all-polymer blends. Horizontal bars are the uncertainties reported with the original measurements, or the spread between two independent determinations where both were reported. For ZR1:Y6 the generation efficiency is normalised to its value at $-2$~V rather than given on an absolute scale. Source data are provided with this paper.}
\label{fig:experiment}
\end{figure}

The numerical verification showed that the $J_{\mathrm{ph}}$--$V_{\mathrm{eff}}$ procedure returns a collection efficiency, modulated by any field dependence of charge generation, rather than a generation yield. To test this against measurement we require systems whose free-charge generation efficiencies are known independently of the current--voltage curve. 
We assemble ten organic solar cells for which this is the case: four polymer:non-fullerene blends we reported previously \cite{TIQE}; a polymer:fullerene blend, PCDTBT:PC$_{70}$BM \cite{Zeiske2022}; an all-small-molecule blend, ZR1:Y6 \cite{Gerber2024}; and four all-polymer blends \cite{Alam2026}. The last six are taken from current--voltage characteristics published by other groups. The $J_{\mathrm{ph}}$--$V_{\mathrm{eff}}$ construction requires $V_{0}$, which cannot be located reliably from those published curves, so Figures~\ref{fig:experiment}(a) and (b) show the four blends of Ref.~\cite{TIQE} and all ten are compared at short circuit in Figure~\ref{fig:experiment}(c). 

The light current--voltage characteristics of these four blends are shown in Figure~\ref{fig:experiment}(a); their free-charge generation efficiencies were determined using temperature-dependent internal quantum efficiency measurements conducted at short circuit \cite{TIQE}. These include PM6:Y6 and PM6:BTP-eC9, which have near-unity, field-independent generation, $P_{\mathrm{gen}}=0.984\pm0.003$ and $0.993\pm0.003$, respectively, as well as PM6:ITIC and PBDB-T:EH-IDTBR, which carry substantial geminate recombination losses with $P_{\mathrm{gen}}=0.857\pm0.019$ and $0.633\pm0.026$, respectively. The quoted uncertainties are those reported with the original measurements \cite{TIQE}. The corresponding $J_{\mathrm{ph}}$--$V_{\mathrm{eff}}$ curves are shown in Figure~\ref{fig:experiment}(b), where each curve has been normalised to its reverse-bias saturation current $J_{\mathrm{sat,exp}}$. The apparent exciton dissociation efficiencies read from the short-circuit point ($V_{\mathrm{eff}}=V_{0}$) are marked on each curve; these points are not all aligned at saturation. Those of the high-fill-factor blends PM6:BTP-eC9 and PM6:Y6 have risen close to saturation by short circuit, whereas those of PM6:ITIC and PBDB-T:EH-IDTBR are still climbing there, so the normalised photocurrent at short circuit, and hence $P_{\mathrm{diss}}^{\mathrm{app}}$, is reduced by collection losses just as the fill factor is. This is the signature of voltage-dependent collection. The level reached on each curve at short circuit records how completely the photogenerated charge has been extracted at the operating field, not how much of it was generated. The $J_{\mathrm{ph}}$--$V_{\mathrm{eff}}$ procedure returns $P_{\mathrm{diss}}^{\mathrm{app}}=0.990$, $0.971$, $0.967$ and $0.917$ for PM6:BTP-eC9, PM6:Y6, PM6:ITIC and PBDB-T:EH-IDTBR, in order of decreasing fill factor.

Figure~\ref{fig:experiment}(c) plots $P_{\mathrm{diss}}^{\mathrm{app}}$ evaluated at short circuit using the $J_{\mathrm{ph}}$--$V_{\mathrm{eff}}$ procedure against the corresponding free-charge generation yield, $P_{\mathrm{gen}}(F_{\mathrm{sc}})$, for all ten systems. The same pattern runs through the whole set. The measured generation yield spans $0.39$ to $0.99$, a factor of two and a half, while $P_{\mathrm{diss}}^{\mathrm{app}}$ returned by the procedure stays within $0.85$--$0.99$. Seven of the ten are overestimates, by between $0.11$ and $0.46$, and among those seven the overestimate grows monotonically as the true yield falls; the remaining three are underestimates of at most $0.030$. The four all-polymer blends illustrate the point most plainly: each returns an apparent dissociation probability between $0.85$ and $0.92$ for a device whose generation efficiency is $0.39$--$0.48$, an overestimate close to a factor of two in every case. A quantity confined to a narrow band near unity while the yield it is supposed to report varies by a factor of two and a half is not reporting that yield.

These four blends are the systems for which the uncertainty on the generation yield is also known, so they allow the size of the discrepancy to be set against experimental error. In PM6:ITIC and PBDB-T:EH-IDTBR $P_{\mathrm{diss}}^{\mathrm{app}}$ exceeds the measured generation yield by $0.11$ and $0.28$, that is by six and eleven times the reported uncertainty. For the other two the deviations are small and of the opposite sign, $-0.003$ for PM6:BTP-eC9 and $-0.013$ for PM6:Y6; the former is within the measurement uncertainty and no significance should be attached to it. Within this subset the evidence therefore rests on the two systems with substantial geminate losses, where the discrepancy is far outside any experimental error; the six literature systems extend the same comparison over a far wider range of generation yields, where the overestimate is larger than any plausible measurement error.

For the systems taken from the literature $P_{\mathrm{diss}}^{\mathrm{app}}$ was evaluated between short circuit and reverse bias, where the dark current is either negligible or, where reported \cite{Alam2026}, subtracted; neglecting it lowers $P_{\mathrm{diss}}^{\mathrm{app}}$, so the values plotted are conservative. The reason for the behaviour seen across the whole set can be traced back to the fact that, in case of field-independent charge generation, $P_{\mathrm{diss}}^{\mathrm{app}}$ [Equation~(\ref{eq:Pdiss})] becomes independent of $P_{\mathrm{gen}}$, yielding the short-circuit collection efficiency and thus tracking the fill factor instead.

\section*{Discussion}
The $J_{\mathrm{ph}}$--$V_{\mathrm{eff}}$ procedure used to evaluate $P_{\mathrm{diss}}^{\mathrm{app}}$ measures neither exciton dissociation nor charge generation efficiency. When this procedure is applied to a device with field-independent charge generation the generation ratio of Equation~(\ref{eq:ratio}) is exactly unity, and what the procedure returns is the short-circuit collection efficiency, a quantity fixed by the same extraction-to-recombination competition that fixes the fill factor and that therefore carries no information beyond it. It therefore provides no further insight for material systems comparisons. Even in the limit of ideal transport the extracted probability does not reach unity but saturates at a first-order ceiling set by either recombination with the injected charge or diffusion-induced surface recombination. In this case, the reported dissociation deficit of a few percent is generic and unrelated to any geminate loss. Measurements on a wide range of organic solar cell systems with independently known free-charge generation yields confirm that the apparent exciton dissociation efficiency tracks the collection rather than the generation.

The practical consequence is that a reported improvement in $P_{\mathrm{diss}}^{\mathrm{app}}$ following a change of material or processing conditions need not indicate improved charge generation. The quantity conflates generation with collection, and whether the true generation yield is over- or underestimated depends on the transport quality and on the yield itself, so the direction of the error cannot be read off the reported number. Consequently, comparisons of exciton dissociation efficiency drawn from the $J_{\mathrm{ph}}$--$V_{\mathrm{eff}}$ procedure are dominated by differences in fill factor. Quantifying generation requires a probe that determines the free-charge generation current ($J_{G}$) independently of the steady-state current--voltage curve, after which the physically-meaningful collection efficiency is $J_{\mathrm{sc}}/J_{G}$. Techniques such as time-delayed collection-field and temperature-dependent internal quantum efficiency measurements provide such probes \cite{Kniepert2011,TIQE}. However, neither is entirely free of artefacts: non-geminate recombination during extraction can introduce an apparent field dependence into time-delayed collection-field transients, and the conditions under which this matters, together with protocols for minimising it, have been examined in detail \cite{Wurfel2018,Gerber2024}; temperature-dependent internal quantum efficiency measurements become increasingly transport-limited as the temperature is lowered \cite{TIQE}. These are, however, quantifiable corrections to methods that do measure generation, whereas the $J_{\mathrm{ph}}$--$V_{\mathrm{eff}}$ procedure does not measure it at all. It remains the case that it is against an independently-measured $J_{G}$, and not against the saturated photocurrent, that collection should be normalised.

Two model simplifications should be noted. First, the 400~nm device assumes Beer--Lambert generation, whereas real devices exhibit wavelength-dependent optical interference. Figure~\ref{fig:ceiling}(d) nevertheless establishes that $P_{\mathrm{diss}}^{\mathrm{app}}$ depends on the spatial generation profile and illumination direction. Interference modifies the profile, not the underlying dependence. Second, the contact model assumes complete loss of minority carriers reaching the wrong electrode. Selective transport layers suppress this loss, reducing surface recombination while a first-order loss still remains. Neither simplification affects Equation~(\ref{eq:ratio}), which relates collection efficiencies independently of the mechanisms governing them. Moreover, all devices in Figure~\ref{fig:experiment} incorporate selective transport layers and still exhibit apparent dissociation probabilities far exceeding their measured generation yields.

\section*{Methods}

\subsection*{Analysis of published current--voltage data}

The current--voltage characteristics of the four polymer:non-fullerene blends, together with the independently determined free-charge generation efficiencies against which the outcome of the $J_{\mathrm{ph}}$--$V_{\mathrm{eff}}$ procedure is compared, are taken from Reference~\cite{TIQE}. The four blends are PM6:Y6, PM6:BTP-eC9, PM6:ITIC and
PBDB-T:EH-IDTBR, with active layer thicknesses of $100$~nm in all cases except
PBDB-T:EH-IDTBR, for which $d=90$~nm. The generation efficiencies quoted in the
text, $P_{\mathrm{gen}}=0.984\pm0.003$, $0.993\pm0.003$, $0.857\pm0.019$ and $0.633\pm0.026$ respectively, were
obtained in that work by temperature-dependent internal quantum efficiency
spectroscopy performed at short circuit, and therefore already correspond to the
short-circuit field $F_{\mathrm{sc}}$. They may consequently be compared directly
with $P_{\mathrm{diss}}^{\mathrm{app}}$ evaluated at short circuit through
Equation~(\ref{eq:Pdiss}) without further field correction. The six further systems compared in Figure~\ref{fig:experiment}(c) are taken from the published current--voltage characteristics and generation efficiencies of References~\cite{Zeiske2022}, \cite{Gerber2024} and \cite{Alam2026}: PCDTBT:PC$_{70}$BM from Figure~4 of Reference~\cite{Zeiske2022}, ZR1:Y6 from Figures~6(b) and 6(d) of Reference~\cite{Gerber2024}, and the four all-polymer blends from Figure~6 of Reference~\cite{Alam2026}. For these $P_{\mathrm{diss}}^{\mathrm{app}}$ was evaluated between short circuit and reverse bias, the dark current being negligible there or, where reported \cite{Alam2026}, subtracted. The
$J_{\mathrm{ph}}$--$V_{\mathrm{eff}}$ procedure was applied to these data exactly
as described for the simulations below, with $J_{\mathrm{sat,exp}}$ read at the
most negative bias reached in each measured sweep.

\subsection*{Numerical drift-diffusion simulations}
Light and dark current--voltage characteristics were computed with a
one-dimensional drift-diffusion model of a single intrinsic active layer between
two contacts. The model solves Poisson's equation,
$\partial_{x}(\varepsilon_{\mathrm{r}}\varepsilon_{0}\partial_{x}\psi)=q(n-p)$, together with
the steady-state continuity equations for electrons and holes,
$\partial_{x}J_{n}/q=R-G$ and $-\partial_{x}J_{p}/q=R-G$, with drift-diffusion
fluxes $J_{n}=q\mu_{n}(-n\partial_{x}\psi+ (k_{\mathrm{B}}T/q)\partial_{x}n)$ and
$J_{p}=q\mu_{p}(-p\partial_{x}\psi-(k_{\mathrm{B}}T/q)\partial_{x}p)$, where $\psi$ is the
electrostatic potential. The equations are discretised on a uniform grid of
$171$ nodes ($221$ for the $400$~nm device) using the Scharfetter--Gummel
scheme, which keeps the carrier flux exponentially accurate across a mesh
element at high field \cite{Scharfetter1969,selberherr1984analysis}, and the
resulting system is solved by Gummel iteration: Poisson's equation and the two
continuity equations are updated in turn until the potential and both carrier
densities change by less than one part in $10^{6}$ between successive sweeps.
Both contacts are Ohmic in the sense that $np=n_{\mathrm{int}}^{2}$ is imposed at
the boundary nodes, with the equilibrium densities there fixed by the injection
barrier $\Phi$; a minority carrier reaching the wrong electrode is therefore
lost at once, which is the limit of infinite surface recombination velocity and
hence the most pessimistic case for collection. An implementation of this model is
described in Reference~\cite{kay2024new}. Excitons are generated uniformly throughout
the active layer at a rate $G_{\mathrm{ex}}$, and unless stated otherwise the only
bulk loss channel available to the resulting free carriers is bimolecular
recombination at a rate
\begin{equation}\label{eq:BiRec}
R=\gamma\,\beta_{\mathrm{L}}\,(np-n_{\mathrm{int}}^{2}),\qquad
\beta_{\mathrm{L}}=\frac{q(\mu_{n}+\mu_{p})}{\varepsilon_{\mathrm{r}}\varepsilon_{0}},
\end{equation}
where $n$ and $p$ are the electron and hole densities, $\beta_{\mathrm{L}}$ is the
Langevin coefficient, $\mu_{n}$ and $\mu_{p}$ are the free electron and hole
mobilities, $n_{\mathrm{int}}$ is the intrinsic carrier density, and $\gamma$ is a
reduction factor scaling the recombination strength. The reduction factor is treated phenomenologically; its microscopic origin, which in donor:acceptor blends is usually attributed to re-dissociation of charge-transfer states, is outside the scope of this work. The conclusions drawn here depend only on the collection efficiency that results, and not on the mechanism that sets it, since the ratio in Equation~(\ref{eq:Pdiss}) returns $P_{\mathrm{coll}}$ whatever process determines it. Unless stated otherwise the
simulations used $T=300$~K, an active layer thickness $d=100$~nm, a band gap
$E_{\mathrm{g}}=1.4$~eV, symmetric injection barriers
$\Phi=(E_{\mathrm{g}}-qV_{\mathrm{bi}})/2=0.25$~eV at both contacts giving a built-in
voltage $V_{\mathrm{bi}}=0.9$~V, balanced mobilities
$\mu_{n}=\mu_{p}=2\times10^{-4}~\mathrm{cm^{2}\,V^{-1}\,s^{-1}}$, a relative
permittivity $\varepsilon_{\mathrm{r}}=3.5$, an effective density of states of
$10^{20}~\mathrm{cm^{-3}}$ for both carriers, and an exciton generation rate
$G_{\mathrm{ex}}=10^{22}~\mathrm{cm^{-3}\,s^{-1}}$, which yields a saturated
current $|J_{\mathrm{sat}}|=qG_{\mathrm{ex}}d\simeq16~\mathrm{mA\,cm^{-2}}$ that is comparable to one-sun operation.

Trap-assisted recombination was included in Figure~\ref{fig:ceiling}(a) through a Shockley--Read--Hall term, $R_{\mathrm{SRH}}=c_{n}c_{p}N_{\mathrm{t}}(np-n_{\mathrm{int}}^{2})/[c_{n}(n+n_{1})+c_{p}(p+p_{1})]$, with midgap traps ($n_{1}=p_{1}=n_{\mathrm{int}}$) at $N_{\mathrm{t}}=10^{16}~\mathrm{cm^{-3}}$ and capture coefficients $c_{n,p}=q\mu_{n,p}/\varepsilon_{\mathrm{r}}\varepsilon_{0}$ taken from the Langevin encounter rate, so that $c_{n}+c_{p}=\beta_{\mathrm{L}}$. Because this term shares the factor $np-n_{\mathrm{int}}^{2}$ with Equation~(\ref{eq:BiRec}), it was implemented as a position-dependent effective reduction factor updated at each Gummel iteration; setting $N_{\mathrm{t}}=0$ reproduces the trap-free results.

Free-charge generation was prescribed rather than inferred. In the first set of
simulations the generation efficiency was set to unity at every field, so that
the local generation rate is $G=G_{\mathrm{ex}}$ throughout and any deficit
returned by the procedure must be a collection loss. Field-dependent generation
was then introduced by scaling the local rate as
$G(F)=G_{\mathrm{ex}}P_{\mathrm{gen}}(F)$ with the Onsager--Braun yield
\begin{equation}\label{eq:OnsagerBraunGen}
P_{\mathrm{gen}}(F)=\frac{P_{\mathrm{gen}}^0\,f(F)}{P_{\mathrm{gen}}^0\,f(F)+1-P_{\mathrm{gen}}^0};\qquad
f(F)=\frac{I_{1}\!\left(2\sqrt{2b_F}\,\right)}{\sqrt{2b_F}},
\end{equation}
in which $I_{1}(u)$ is the modified Bessel function of the first kind of order
one, $b_F=q^{3}|F|/[8\pi\varepsilon_{\mathrm{r}}\varepsilon_{0}(k_{\mathrm{B}}T)^{2}]$,
$\varepsilon_0$ is the vacuum permittivity, $k_{\mathrm{B}}$ is Boltzmann's
constant, and $P_{\mathrm{gen}}^0$ is the zero-field yield. A local field
$|F|=(V_{\mathrm{bi}}-V)/d$ was assumed throughout the device. Equation
(\ref{eq:OnsagerBraunGen}) is equivalent to the conventional form
$f(F)=J_{1}(2\sqrt{-2b_F})/\sqrt{-2b_F}$ of the Onsager--Braun model
\cite{Onsager1938,Braun1984} through the identity
$I_{\alpha}(x)=i^{-\alpha}J_{\alpha}(ix)$, and expanding the Bessel series
\cite{riley2011essential} gives
$f(F)=1+b_F+b_F^{2}/3+b_F^{3}/18+b_F^{4}/180+\dots$, such that $f(0)\to1$ and
$P_{\mathrm{gen}}(0)=P_{\mathrm{gen}}^{0}$ in the zero-field limit. The zero-field yields
$P_{\mathrm{gen}}^0=1$, $0.8$ and $0.6$ used in Figure~\ref{fig:recipe}(b) fix the
short-circuit ceilings $P_{\mathrm{gen}}(F_{\mathrm{sc}})=1.00$, $0.97$ and $0.92$
quoted in the text.

For each parameter set the full light and dark sweeps were computed from high
reverse to high forward bias, the photocurrent was evaluated from
Equation~(\ref{eq:Jph}), $V_{0}$ was located as the voltage at which
$J_{\mathrm{ph}}=0$, and $J_{\mathrm{sat,exp}}$ was read at high reverse bias. The apparent dissociation efficiency was then determined using Equation~(\ref{eq:Pdiss}), 
exactly the same way an experimentalist would use to obtain it. For the sweeps of
Figure~\ref{fig:recipe}(c) and (d) the mobilities were varied from
$5\times10^{-5}$ to $2\times10^{-3}~\mathrm{cm^{2}\,V^{-1}\,s^{-1}}$ and the
reduction factor from $\gamma=0.05$ to $1$ at fixed $V_{\mathrm{bi}}$ and $d$,
while for Figure~\ref{fig:ceiling} the injection barrier and illumination
intensity were varied as indicated in the caption. For Figure~\ref{fig:recipe}(f) the generation yield was held constant and field independent, and the mobility was varied at $\gamma=1$ so as to span $\alpha=0.1$ to $100$, with $J_{\mathrm{sat,exp}}$ read at $-4$~V. The $400$~nm sweeps of Figure~\ref{fig:recipe}(c), (d) and (e) used mobilities from $10^{-3}$ to $3\times10^{-2}~\mathrm{cm^{2}\,V^{-1}\,s^{-1}}$ and $\gamma=0.05$ to $1$, with the exciton generation rate scaled to $2.5\times10^{21}~\mathrm{cm^{-3}\,s^{-1}}$ so that $qG_{\mathrm{ex}}d$ matched the $100$~nm reference. Since $\alpha^{2}$ depends on the mobility and the reduction factor only through the ratio $\gamma/\mu$, the same map is obtained for any other combination giving the same $\alpha$.

\section*{Data availability}
All data underlying the figures of this study are provided in the Source Data
file supplied with this paper, which contains the numerical values plotted in
every panel of Figures~\ref{fig:recipe}, \ref{fig:ceiling} and
\ref{fig:experiment}, one sheet per panel. Figure~\ref{fig:StateDiagram} is a
schematic and has no underlying data. The measured current--voltage
characteristics of PM6:Y6, PM6:BTP-eC9, PM6:ITIC and PBDB-T:EH-IDTBR, and the
free-charge generation efficiencies determined for them, were reported in
Reference~\cite{TIQE}; the remaining six systems are taken from the published
figures of References~\cite{Zeiske2022}, \cite{Gerber2024} and \cite{Alam2026},
as detailed in Methods, and the values read from them are tabulated in the
Source Data file. The simulated datasets are generated by the deposited code
from the parameters given in Methods.

\section*{Code availability}
The drift-diffusion model used to generate the simulated characteristics in
this work, together with worked examples that reproduce the analysis, is
available at Zenodo under
\url{https://doi.org/10.5281/zenodo.23389108}. An implementation of the solver
is also described in Reference~\cite{kay2024new}.

\bibliographystyle{unsrt}
\bibliography{jph_veff}

\section*{Acknowledgements}
At Swansea University's Centre for Integrative Semiconductor Materials (CISM), P.M. is the Director, A.A. holds an honorary position, D.B.R. is a Research Officer, and A.M.K. is a Post-Doctoral Research Assistant.

\section*{Funding}
CISM was funded by the UK Research Partnership Investment Fund (UKRPIF) through Research England. P.M. is an EPSRC Fellow funded under project MANTISS, Molecular Absorbers for Novel Transducing Imaging and Sensing Semiconductor Photodetectors (UKRI4483). O.J.S. acknowledges funding from the Research Council of Finland through Project No. 357196.

\section*{Author contributions}
A.A. and O.J.S. conceptualised the work. A.A. performed the simulations and drafted the manuscript. P.M. oversaw the project. A.M.K. developed the computational framework under the supervision of O.J.S. and A.A. D.B.R. contributed to the interpretation of the results. All authors contributed to the development of the manuscript.

\section*{Competing interests}
The authors declare no competing financial or non-financial interests.

\end{document}